\documentclass[fleqn,12pt,]{article}
\usepackage{hyperref}

\usepackage[utf8]{inputenc}
\usepackage{comment}
\usepackage{amsthm}
\usepackage{tabularray}
\usepackage{threeparttable}
\usepackage[bottom]{footmisc}
\usepackage[papersize={8.3in,11.7in},vmargin={1.0in,1.0in}, hmargin={1.0in,1.0in}]{geometry}
\usepackage{times}
\usepackage{booktabs}

\usepackage[round]{natbib}

\usepackage{authblk}
\usepackage{soul}
\usepackage{verbatim, booktabs, multirow, hyperref,dirtytalk}
\usepackage{graphicx, rotating}
\usepackage[utf8]{inputenc}
\usepackage{enumerate}
\usepackage{euscript}
\usepackage{bbm}
\usepackage{ulem}
\usepackage{amsmath, amsthm,amsfonts, amssymb}

\usepackage{extarrows}
\usepackage{tikz}
\usetikzlibrary{calc}
\hypersetup{colorlinks,linkcolor={blue},citecolor={blue},urlcolor={red}}
\usetikzlibrary{positioning}
\usetikzlibrary{arrows.meta}
\usepackage{caption}
\usepackage{subcaption}
\usepackage{booktabs}

\usepackage{pgfplots}
\usepgfplotslibrary{dateplot}
\usepackage{pdfpages}
\usepackage{bm}
\usepackage{float}

\newtheorem*{rmk}{Remark}
\newtheorem{lemma}{Lemma}

\newtheorem{defn}{Definition}

\newtheorem{example}{Example}
\newtheorem*{example*}{Example}
\newtheorem{proposition}{Proposition}

\usepackage{setspace}

\usepackage{verbatim, booktabs, multirow, hyperref}
\usepackage{bbm}
\usepackage{lmodern}
\usetikzlibrary{intersections}

\usepackage{tikz}
\usetikzlibrary{positioning}
\usepackage{tikz-3dplot}
\usepackage{soul}
\usetikzlibrary{arrows, patterns, decorations.pathmorphing,backgrounds,positioning,fit,petri,trees}
\usepackage{pgfplots}
\usepackage{pgfplotstable}

\usetikzlibrary{arrows}
\usetikzlibrary{calc}
\usetikzlibrary{positioning}
\usetikzlibrary{arrows.meta}
\usetikzlibrary{decorations.pathreplacing,calligraphy}
\usetikzlibrary{shapes.geometric}

\pgfplotsset{compat=1.18}
\usepackage{enumitem}

\title{Evolving Rules: Imitation and Best Response Learning in Cournot Oligopoly \footnote{We are indebted to Dan Friedman and J\"org Oechssler for helpful comments. This work was conducted while Xiaomeng Ding was visiting the Department of Economics at University of Essex, whose hospitality is greatly acknowledged. Ding acknowledges financial support of the China Scholarship Council program (No. 202406040148). Zhang acknowledges financial support from the National Natural Science Foundation of China (No. 72131003), the Beijing Natural Science Foundation (No. Z220001), and the National Key Research and Development Program of China (No. 2020YFA0712900). All authors contributed equally.}}

\author{Xiaomeng Ding\footnote{Laboratory of Mathematics and Complex Systems, Ministry of Education, School of Mathematical Sciences,
Beijing Normal University, 100875 Beijing, P. R. China. Email: dingxm0616@gmail.com. }\quad\quad Simon Weidenholzer\footnote{Department of Economics, University of Essex, CO4 3SQ, Colchester, UK.}
\quad\quad Boyu Zhang\footnote{Laboratory of Mathematics and Complex Systems, Ministry of Education, School of Mathematical Sciences, Beijing Normal University, 100875 Beijing, P. R. China. Email: zhangby@bnu.edu.cn.}}

\date{\today}

\begin{document}

\maketitle

\begin{abstract}
\vspace{.5cm}
\noindent  
We study evolutionary dynamics in which firms endogenously revise the behavioral rules that govern strategy revisions in symmetric Cournot oligopoly. Specifically, we consider two principles that guide rule revision, \textit{No-Birth} and \textit{Survival-of-the-Fittest}, both grounded in imitation-based heuristics. We show that, under these principles, all firms eventually adopt the same behavioral rule. Focusing on two classical rules, myopic best response and imitation, we demonstrate that rule revision plays a crucial role in determining long-run equilibria in Cournot oligopoly. The set of long-run equilibria includes the state where all players use best response learning and choose the Nash equilibrium quantities and states where all firms use imitation learning and choose specific symmetric quantities which include (but are not necessarily restricted to) Walrasian quantities. 
Our results extend to more general aggregative games.  

\vspace{.5cm}
\end{abstract}
\setstretch{1.5} 
\thispagestyle{empty}

\section{Introduction}

Often individuals use simple behavioural rules to determine their actions in strategic settings \citep{nax2016learning, huck2017payoff,kovavrik2018learning,newton2018evolutionary}. But different behavioral rules may prescribe different actions and subsequently lead to different societal outcomes. We are consequently interested to understand which rules economic agents come to adopt and what the implications are for action choice when agents choose rules that have historically performed well for themselves and others.\footnote{Rule revision captures more widespread phenomena  than  decision makers occasionally adjusting their behavioral rules. In business contexts, firms or institutional agents often operate with finite managerial tenure and routine leadership turnover, circumstances that naturally create opportunities for reassessing past performance and adjusting decision-making rules. See e.g., \cite{denis1995performance} and \cite{weisbach1995ceo} for empirical studies documenting changes in firms' strategies following managerial turnover. }

We cast our research question in the context of Cournot competition which is the workhorse model of imperfect competition in Industrial Organization but may also more widely be considered as a metaphor for social dilemmas with a conflict between maximizing individual outcomes and achieving cooperative outcomes. In this setting firms can either choose their actions via imitation \citep[e.g.][]{vega1997evolution} or through myopic best response \citep[e.g.][]{fisher1961stability}. Under imitation learning firms simply choose the quantity that has yielded the highest profit in the industry in the previous period, and under best response learning, they play a best response to the previous distribution of quantities. 
When firms consider switching rules, we are interested in scenarios where they are guided by two simple principles: the \textit{No-Birth} principle, which postulates that no new rule may emerge other than those already in use in the industry and the \textit{Survival-of-the-Fittest} principle, which assumes that rules that have earned the highest average historical payoffs to some firm are adopted with positive probability. These two principles capture a broad class of criteria grounded in imitation-based heuristics, where rules with the highest historical payoffs are adopted. Rule adoption, however, is not necessarily confined to these top rules; for example, a rule that outperforms a firm’s current choice may also be selected. 
We augment this adjustment process by allowing for mistakes in action choice and in rule choice. Considering the case when the probability of mistakes vanishes allows us to characterize the set of {\it long-run equilibria}  \citep{kandori1993learning, young1993evolution} as predictions for the long-run behaviour of the dynamics. Following this literature, the set of long-run equilibria can be identified by studying the relative robustness of absorbing sets  to mistakes. 

To appreciate and understand our results, it is useful to consider the benchmark cases where i) all firms use myopic best response  to determine their quantities and ii) all firms use imitation learning. In the first case, \cite{Cournot} has already discussed  convergence of a myopic best response process to the Cournot Nash equilibrium under linear demand and costs.\footnote{Subsequent studies show that myopic best response dynamics also converge to the Nash equilibrium for more general demand and cost functions \citep[Chapter 4]{dubey2006strategic,vives1999oligopoly}. In addition, the convergence result also holds for other belief-based learning rules, e.g., the fictitious play \citep[Chapter 2]{fudenberg1998theory}, the Bayesian learning \citep{kalai1993rational}. } In the second case where  firms imitate quantities that have yielded the highest profit in the previous period, \cite{vega1997evolution} has shown that the Walrasian equilibrium, where firms act as (if they were) price takers, is the unique long-run equilibrium. To understand this, note that the Walrasian quantity has the property that any firm choosing it will enjoy a payoff advantage over firms choosing different quantities. Thus, any mistake towards the Walrasian quantity will be imitated by other firms whereas a single mistake away from the profile where all firms choose the Walrasian quantity will not be followed. Note that the Walrasian equilibrium -underpinned by imitation- is more competitive than the Cournot Nash equilibrium arising under myopic best response learning; the former yields strictly higher output and lower profit than the latter.

In the case where firms may adjust their rules, it is useful to observe that under the \textit{No-Birth-} and the \textit{Survival-of-the-Fittest-} principle the dynamics will converge to states where all firms use the same behavioral rule. If the behavioral rule used by all firms is best response learning, then in fact all firms will have to choose the Nash quantity. In contrast, when all firms use imitation learning, states in which all firms choose the same quantity are absorbing. 

Given this characterization of absorbing sets, we then proceed to identify the number of mistakes required for transitions among these states. At the core of our results is the fact that only one mistake in rule choice is required to move from the state where all firms are imitators and choose the Walrasian quantity to the state where all firms are best responders and choose the Nash quantity. At the same time, it is possible to move from the state where all firms are best responders choosing the Nash quantity to the state where they are all imitators choosing the Nash quantity. Starting from the latter state, we can connect a whole array of states where firms are imitators via different mistakes in action choice. While these action mistakes may be directly towards the Walrasian quantity, they can also be towards other quantities that are likewise relative-payoff improving. Putting these observations together allows us to characterize the set of long-run equilibria. If mistakes in rule choice happen less frequently than in action choice, then we have two types of long-run equilibria: in one, everybody is a best responder and chooses the Nash quantity, and in the other, everybody is an imitator choosing the Walrasian quantity. If, however, action and rule mistakes occur at the same rate, then in addition to the two types above, states where all firms are imitators all choosing the same quantity from an interval are long-run equilibria.\footnote{If there are only two firms, the lower bound of this interval is the Nash quantity and the upper bound is a specific quantity above the Walrasian quantity. If there are more than two firms and the grid of feasible output is sufficiently fine, the lower bound is zero while the upper bound is a specific quantity strictly greater than the Walrasian quantity.}

These results are surprising, especially given that imitators outperform best responders in the long run when the two types of learners coexist \citep{schipper2009imitators}.\footnote{\cite{schipper2009imitators} shows that any Pseudo-Stackelberg state where imitators are weakly better off than best responders is absorbing. Moreover, when mistakes are possible, the Nash equilibrium cannot be the unique prediction and hence the imitators’ long run average per-period payoff is strictly higher than that of best responders.}$^,$\footnote{Further support for the superior performance of imitation learning over other rules is provided by \cite{hehenkamp2008imitators}, \cite{duersch2010rage, duersch2012unbeatable} or \cite{bossan2015evolution}.} The intuition is that imitators, acting as price takers, choose a quantity that yields the highest relative payoff, while best responders behave like Stackelberg followers. Consequently, it is natural to conjecture that, imitation learning drives out best response learning under payoff monotone selection dynamics. Our analysis, however, reveals that best responders can also dominate the population in the long run. 
The key mechanism is that the average historical payoff of imitators choosing the Walrasian equilibrium may (at least temporarily) be dominated by the payoff received by best responders.

Another implication of our results is that in the presence of rule revisions the Walrasian equilibrium is not necessarily the unique long-run outcome when all firms employ imitation learning. Specifically, when action and rule mistakes occur at the same frequency, the Cournot Nash equilibrium (arising from best response learning) serves as an intermediate stability plateau, facilitating deviation from the Walrasian equilibrium and giving rise to a spectrum of output levels that may even include the collusive quantity. This contrasts with the well-established results based on imitation which typically hold learning rules fixed, and highlights the role of endogenous rule revision in shaping long-run outcomes.

\subsection*{Related literature}

Early theoretical work on learning or evolutionary approaches to Cournot competition assumes that all firms adhere to the same behavioral rule, such as myopic best response or imitation learning, with long-run outcomes depending on the rule: the former converges to the Nash equilibrium, whereas the latter yields the Walrasian outcome. However, experimental studies indicate that the coexistence of multiple rules within the same environment, rather than the dominance of a single behavioral rule \citep{huck1999learning,huck2002stability,bosch2003imitation, apesteguia2010imitation,bigoni2013information}. In particular, the coexistence of both best response and imitation learning is frequently observed, with the prevalence of each affected by the information settings. Furthermore, recent evidence from long-horizon experiments suggests that subjects adjust their behavioral rules over time \citep{friedman2015imitation, oechssler2016imitation}, even though the rule adjustment is significantly affected by information feedback \citep{huck2017payoff}. In a related vein, \citet{alos2021multiple} provide evidence that, in Cournot oligopolies, imitation and more deliberative rules such as myopic best response coexist at the intra-individual level, with the  rules used determined by cognitive interaction.  

Departing from the assumption of a single behavioral rule, \cite{schipper2009imitators} studies a theoretical model in which a fixed set of best responders and and a fixed set of imitators repeatedly play a symmetric Cournot oligopoly. The analysis shows that the long-run distribution converges to a subset of those states corresponding to Stackelberg equilibria, where imitators earn a strictly higher average per-period payoff than best responders. Building on this finding, \cite{schipper2009imitators} conjectures that if agents were allowed to revise their rules according to past performance, and if the occurrence of rule revision were sufficiently infrequent relative to quantity revision, the share of imitators would increase and eventually dominate the population. To the best of our knowledge, however, no study has yet provided an explicit theoretical prediction of how the two rules evolve in the context of Cournot oligopolies.\footnote{\citet{thijssen2009approximate} is closely related to our work in terms of motivation. However, the paper only establishes approximate results under the assumption of an extremely low frequency of rule revision and there is no explicit definition of the rule revision procedure.} In our study, we explicitly defines a rule revision protocol that refers to historical average payoffs and provide a complete characterization of the long-run outcomes. Our analysis, however, challenges the conjecture put forward by \citet{schipper2009imitators} that only imitators can be around in the long run.

\citet{juang2002rule} examines rule evolution in a setting where agents are randomly matched into pairs to play $2\times2$ coordination games, with some adopting best response learning and others relying on imitation learning.\footnote{Note that \cite{juang2002rule}’s definition of imitators differs slightly from ours: in his model, imitators select the action that yields the highest average payoff over the population.} When behavioral rules are fixed, the analysis shows that the long-run outcome is determined by the relative proportion of the two rules in the population. In contrast, when agents are allowed to revise their rules according to the \textit{experimental} criterion, they select with positive probability among the rules currently in use in the population without reference to historical performance, and in the long run all agents coordinate at the Pareto-efficient equilibrium, with the prevailing rule being either imitation or best response. We advance \cite{juang2002rule}'s framework by explicitly conceptualizing the incentives underlying rule revision. 

A more recent contribution by \citet{juang2021rules} introduces the \textit{monotonicity} principle, under which agents take the performance of rules into account. While sharing some similarities, our study differs in several important dimensions. First, we restrict attention to symmetric Cournot oligopolies and focus on two classical behavioral rules, best response and imitation, motivated by robust experimental evidence. By contrast, \citet{juang2021rules} consider a random matching framework in which agents, in every period, are randomly matched into groups and assigned random roles to play normal-form games. Besides, a behavioral rule is defined as a mapping from a finite set of histories to the action set so that the number of possible rules grows exponentially with memory length. Second, under our \textit{Survival-of-the-Fittest} principle, each agent evaluates the performance of its current rule relative to others using past average payoffs. In contrast, under their \textit{monotonicity} principle, agents make period-by-period comparisons of their own payoff with that of agents playing the same role across different matches, and only a rule that has outperformed any alternative in every period and match is regarded as attractive. Third, \cite{juang2021rules} assume that mistakes happen only during rule revision. Moreover, these mistakes are payoff-dependent, which is crucial for their main result that only action profiles that maximize the minimum payoff among agents are selected in the long run when the set of feasible rules is sufficiently rich.\footnote{For example, in the Cournot game studied in this paper, the unique action profile that maximizes the minimum payoff among agents is given by the profile where all firms produce the collusive quantity. It is worth noting, however, that the results of \cite{juang2021rules}  do not apply for Cournot games when the set of feasible rules is restricted to  best response and imitation learning (as studied in the present contribution).}

Some studies are conceptually related to ours but methodologically different. \cite{conlisk1980costly} develops a discrete-time population dynamics with two behavioral types. One type pays a fixed cost to adapt directly to the stochastically evolving social state, while the other type incurs no cost but can only follow a lagged social convention, thereby suffering losses from its deviation relative to the true state. In this model, the performance of behavioral rules is assessed by the average per-agent payoffs of the two types in the previous period, which monotonically determine their respective population shares. \cite{droste2002endogenous} considers an infinite population of firms randomly matched each period to play Cournot duopolies. Firms choose from a finite set of behavioral rules in each period, with population shares evolving according to a replicator equation with mutation noise, where realized profits from the previous period serve as the measure of fitness. Their results show that the coevolution of population and quantity dynamics in Cournot oligopolies can be complex and endogenous fluctuations may emerge in the long run. Beyond methodological differences, the main distinction from our study is that, in these studies, rule shares are updated each period, leaving no scope for assessing rule performance over longer horizons.

The rest of this paper is organized as follows. Section \ref{section: the model} introduces the model and the dynamics of action and rule revision. Section \ref{section: results} conducts a long-run analysis and presents the results. Section \ref{section: conclusion} concludes. All proofs are relegated to Appendix \ref{App: A}. Appendix \ref{App: B} pursues the extension to aggregative games.

\section{The model}
\label{section: the model}
\subsection{The base game}

We are primarily interested in symmetric Cournot games where $n$ firms produce a homogeneous good.\footnote{Appendix \ref{App: B} generalizes our analysis to aggregative games.} Denote by $N=\{1, \ldots, n\}$ the set of firms and assume $n\geq 2$. Letting  $q_i$ denote the output of firm $i$ and $Q := \sum_{i \in N} q_i$ denote the total industry output, the profit of firm $i$ is given by $\pi(q_i,Q)=p(Q)q_i - c(q_i)$ where $p: \mathbb{R}_{+} \to \mathbb{R}_{+}$ represents the inverse demand function and  $c: \mathbb{R}_{+} \to \mathbb{R}_{+}$ the cost function.

To facilitate our analysis, we impose standard assumptions common in Cournot oligopoly \citep[see, e.g., ][Chapter 5]{tirole1988theory}. Specifically, the inverse demand function $p$ is assumed to be twice differentiable on the closed interval $[0, Q_{\max}]$, where $p(0) > 0$ and $p(Q) = 0$ for all $Q \geq Q_{\max}$. Furthermore, $p$ is strictly decreasing, $p'(\cdot) < 0$, and weakly concave, $p''(\cdot) \leq 0$ on $[0, Q_{\max}]$. Moreover, the cost function $c$ is assumed to be twice differentiable on $\mathbb{R}_{+}$, strictly increasing, $c'(\cdot) > 0$, and weakly convex, $c''(\cdot) \geq 0$. To ensure a nontrivial market, we assume $c^\prime(0)<p(0)$. The best response correspondence of firm $i$ gives the set of profit maximizing quantities when the combined output of the other firms is given by $Q_{-i}$ and is denoted by
\begin{align*}
b_i(Q_{-i}) = \arg\max_{q \in \mathbb{R}_{+}} \pi(q, q+Q_{-i}).
\end{align*}
Note that under our assumptions, the profit function is single peaked so that $b_i$ is a function.

The following two quantities will be of special interest.

\begin{defn}
    The Cournot Nash quantity $q^{N}$ satisfies $p(nq^{N})q^{N}-c(q^{N})\geq p((n-1)q^{N}+q)q-c(q)$ for all $q\in\mathbb{R}_{+}$. 
\end{defn}

\begin{defn}
   The Walrasian quantity $q^{W}$ satisfies $p(nq^{W})q^{W}-c(q^{W})\geq p(nq^{W})q-c(q)$ for all $q\in\mathbb{R}_{+}$. 
\end{defn}

Under the above assumptions, the existence and uniqueness of a strictly positive Cournot Nash quantity is well established \citep{tirole1988theory}. The Walrasian quantity is also strictly positive and unique  \citep{vega1997evolution}. Moreover we have $q^{W}>q^{N}>0$. We denote the profits made when everybody chooses the Cournot Nash output by $\pi^C=\pi(q^C,nq^C)$ and the profits when everybody chooses the Walrasian output by $\pi^W=\pi(q^W,nq^W)$. 

For technical reasons, we assume that each firm selects quantity from a common finite strategy set $\Gamma=\{0, \lambda, 2\lambda, \cdots, \nu\lambda\}$ as its output, where $\lambda>0$ is small and $\nu\in\mathbb{N}$ is large. We further impose that this grid includes $q^{N}$ and $q^{W}$.

\subsection{The dynamic process}
\begin{figure}[h]
    \centering     \includegraphics[width=1\linewidth]{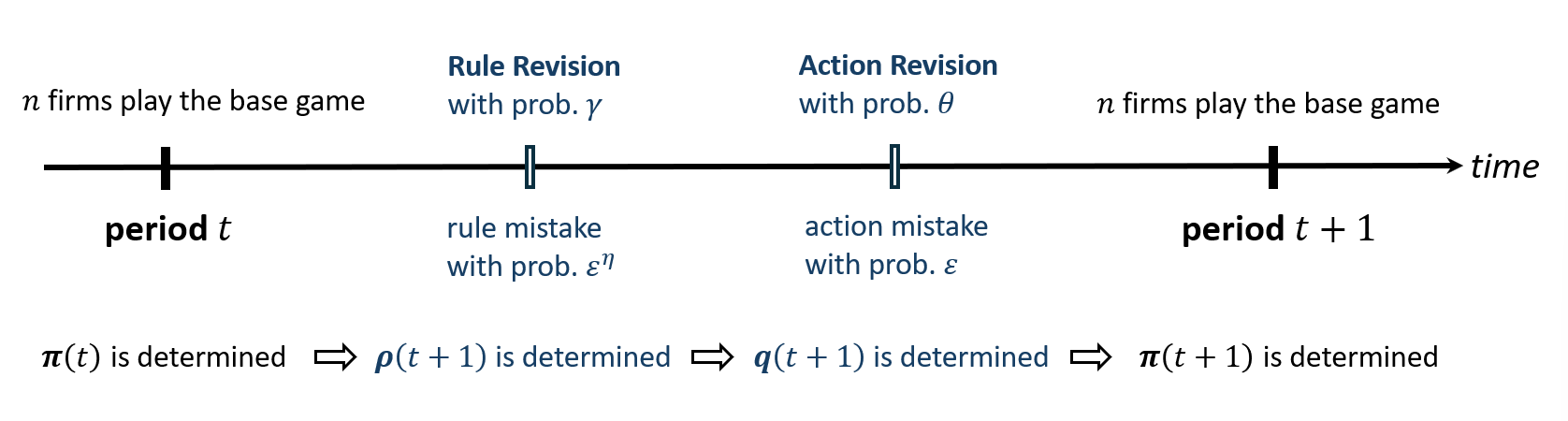}
    \caption{Illustration of the perturbed dynamics with action and rule revision.}
    \label{fig: procedure}
    \vspace{5pt}
\end{figure}

We are interested in a dynamic process where  firms periodically i) use simple behavioral rules to determine their output in the Cournot oligopoly and ii)  decide which rule to use. We proceed to outline the details of this process. Figure \ref{fig: procedure} provides a sketch.

Time is discrete, $t=1, 2, \dots$. At the end of each period $t$, first, with probability $\gamma \in (0,1)$ firm $i$ receives opportunity to revise its rule used for action choice. Such rule revision opportunities are assumed to be independent of firms and time. When receiving a rule revision opportunity, firm $i$ applies a rule selection criterion to decide on its future rule for action revision (specified below), denoted by $\rho(t+1)$. With probability $1-\gamma$ the firm does not receive the opportunity to revise its rule, in which case it sticks to its current rule.   
Note that the expected number of periods between two successive rule revisions for each firm is $ 1/\gamma$.

Following this, with probability $\theta \in (0,1)$, firm $i$ receives opportunity (independent across firms, time and rule revision opportunities) to update its output. Upon receiving such an action revision opportunity, firm $i$ follows its behavioral rule, $\rho_{i}(t+1)$, that prescribes an output for the next period. With the remaining probability $1-\theta$, firm $i$ does not receive the opportunity to revise and retains its current output. 

Our process thus exhibits {\it inertia} in rule revision and in quantity revision.  We remark that inertia in quantity revision\footnote{See also  \cite{huck2002stability} documenting the presence of inertia in quantity revision in experimental Cournot markets.} is required to rule out certain cyclical behaviors\footnote{See e.g., \cite{theocharis1960stability}.} and that our main results also hold under alternative assumptions on strategy revision opportunities.\footnote{For instance, one may assume that a firm  uses  a revised rule immediately to choose its output or that it always adjusts either its rule or its output, but not both in the same period.}

\subsubsection{Action revision}

When revising output, firms do so following behavioral rules. We denote the output profile of firms in period $t$ by $\bm q(t)=(q_1(t), q_2(t),\cdots, q_n(t))\in \Gamma^{n}$, and define $Q(t)=\sum_{i \in N} q_i(t)$. The corresponding profit profile is  $\bm \pi(t)=(\pi_1(t), \pi_2(t),\cdots, \pi_n(t))\in\mathbb{R}^{n}$, where $\pi_i(t)=\pi(q_i(t),Q(t))$ for all $i\in N$. Let $\mathcal{R}$ denote the set of behavioral rules available to firms and  denote by $\bm \rho(t)=(\rho_1(t), \rho_2(t),\cdots, \rho_n(t))\in \mathcal{R}^n$ the profile of rules adopted by firms in period $t$. While our framework can accommodate more general settings,  our analysis focuses on a scenario where firms determine their output through either (myopic) best response learning or through imitation. Letting  $BR$ denote best response learning  and $IM$ imitation learning, we have $\mathcal{R} = \{BR, IM\}$.

When a firm employs $BR$, it chooses a quantity that maximizes its profit given the output of other firms in the previous period.\footnote{See e.g.\ \cite{ellison1993learning,ellison2000basins}, \cite{blume1993statistical}, and \cite{oyama2015sampling}.} Formally, if $\rho_i(t+1) = BR$, 
\begin{align*}
q_i(t+1) \in  \arg\max_{q\in\Gamma} \pi_i(q, q+Q_{-i}(t)),
\end{align*}
where $Q_{-i}(t):=\sum_{j\in N\setminus\{i\}}q_j(t)$ is the combined output of other firms in period $t$. If there are multiple quantities in this set, the firm is assumed to choose all of its best replies with positive probability.\footnote{Note that under our assumption the best response function is single peaked, so that in combination with our finite quantity grid there can be at most two best response quantities.}

When a firm follows $IM$, it chooses a quantity that has earned the highest profit in the previous period among all firms in the industry.\footnote{This rule is known as the \textit{Imitate-the-Best-Max} rule, see e.g.\ \cite{vega1997evolution}, \cite{alos2006imitation, alos2008contagion}, or \cite{apesteguia2010imitation}. We remark that most of our results would go through when considering a larger class of imitative rules such as in e.g. \cite{apesteguia2007imitation} and \cite{alos2014imitation}.} Formally, if $\rho_i(t+1) = IM$, then
\begin{align*}
q_i(t+1) = q_j(t) \quad \text{with}\quad j \in \arg\max_{l \in N} \pi_l(t).
\end{align*}
As before, the firm puts positive probability in all quantities in this set.

\subsubsection{Rule revision}

When firms decide on which behavioral rule to use, each is guided by general principles that capture a set of criteria. In broad terms, firms look back at the historical performance of rules and then choose well-performing rules.

For each firm $i$, let $\tau_{i}(t)\in \mathbb{N}$ be the number of periods it has used its current behavioral rule (since its last rule revision) looking back from period $t$. For technical reasons we have to truncate the number of periods firms look back.\footnote{Specifically, we have to ensure that the state space is finite.} To this end, assume that firms take at most $M$ previous periods into account and note that $M$ can be chosen arbitrarily large (but finite). Denote by $\tilde \tau_{i}(t)=\min\{\tau_{i}(t),M\}$ the number of periods that firm $i$ effectively  takes into consideration. 

In particular, each firm $i$ tracks the average performance of its current rule since its last rule revision by 
$f_i(t)$, which is defined as the average profit\footnote{One may instead consider discounted average profits. In this case, let $\delta \in (0,1]$ be the discount factor and define
\[
f_i(t) = \frac{\sum\limits_{r = t-\tilde\tau_i(t) + 1}^{t} \delta^{t - r} \, \pi_i(r)}{\sum\limits_{r = t-\tilde\tau_i(t) + 1}^{t} \delta^{t - r}}.
\]
All results in the paper remain valid under this alternative assumption.} a firm obtains during the past $\tilde{\tau}(t)$ periods. Formally, 
$$f_i(t) = \frac{\sum\limits_{r= t-\tilde \tau_i(t) + 1}^{t} \pi_i(r)}{\tilde \tau_i(t)}.  $$

With this definition at hand, we define a general class of criteria applied for rule revision. Specifically, we are interested in criteria fulfilling the following two intuitive principles.

\noindent
\textbf{\textit{No-Birth} 
 (\textit{NB})}. A revising firm 
$i$ may adopt a rule in period 
$t+1$ only if that rule is already in use by at least one firm in the industry in period $t$.
Formally, if $\rho_{i}(t+1)=\rho$, $\rho\in\mathcal{R}$, then there exists $j\in N$ such that $\rho_{j}(t)=\rho$.  \\

\noindent
\textbf{\textit{Survival-of-the-Fittest} (\textit{SF})}. A revising firm $i$ switches with positive probability to any of the rules employed by the firms attaining the highest average profits. Formally, $\rho_i(t+1) = \rho_j(t)$ with positive probability if $j\in \arg\max_{l\in N} f_l(t) $.\\

The behavioral rationale underlying these two principles is inherently imitative and embodies the customary monotonicity considerations contemplated
by evolutionary theory, while rules not adopted by any firm are eliminated from consideration and do not reemerge in future revisions.\footnote{The \textit{No-Birth} principle corresponds to the invariance property of many deterministic evolutionary dynamics (e.g. replicator dynamics), whereby once a strategy dies out there is no way to revive it, and hence each face of the state space (a simplex) is invariant \citep{hofbauer1998evolutionary}. In finite populations, related properties are implied by Moran- or Pairwise-Comparison- processes \citep{wang2023imitation}. \cite{juang2021rules} also assume the \textit{No-Birth} principle.}

To illustrate our two principles, we now showcase several examples.

\begin{example}
Under the \textit{Imitate-the-Best-Max} criterion a revising firm employs a behavioral rule that has yielded the highest average profit in the industry. Formally, firm $i$ uses
\[
\rho_i(t+1) = \rho_j(t) \quad \text{with} \quad j\in \arg\max_{l\in N} f_l(t),
\]
randomizing in case multiple rules yield the maximum average profit. Note that rule revision in this contribution is analogous to the \textit{Imitate-the-Best-Max} rule which is discussed in the context of action revision above. Both \textit{NB} and \textit{SF} are satisfied. 
\end{example}

\begin{example} Under the \textit{Imitate-the-Best-Max with Sampling} criterion (see also \citealt{alos2008contagion} for a similar rule in a spatial context)
a firm observes a subset of firms in the industry when revising its rule, with all firms having an equal probability of being sampled. 
Let $m_i(t)$ with $\emptyset \subsetneq m_i(t) \subseteq N$ denote the information sample of firm $i$ after the Cournot competition in period $t$. 
A revising firm then adopts the behavioral rule that has yielded the highest average profit within its information sample. 
Formally, firm $i$ uses
\[
\rho_i(t+1) = \rho_j(t) \quad \text{with} \quad j \in \arg\max_{l \in m_i(t)} f_l(t),
\]
randomizing if multiple rules yield the maximum average profit in the information sample. This criterion satisfies the \textit{NB} and \textit{SF} principles.

A special case of this criterion is obtained when a firm observes only itself and one randomly selected peer, which corresponds to the Pairwise-Comparison rule under strong selection \citep{TraulsenPachecoNowak2007}.
\end{example}

\begin{example}
Under the \textit{Experimental} criterion, as introduced by \cite{juang2002rule}, firms randomly select from the rules that are in use. Formally, a revising firm $i$ uses $\rho_i(t+1) = \rho_j(t)$ with $j\in N$. Again, this criterion satisfies both \textit{NB} and \textit{SF}.
\end{example}

\begin{example}
Under the \textit{Imitate-if-Better} criterion discussed in \cite{duersch2012unbeatable}, a revising firm $i$ keeps its current rule $\rho_{i}(t)$ unchanged if its average profit is maximal in the industry. Otherwise, it switches to a rule that is used by one of the firms with strictly higher average profit. Ties break randomly. Formally,
\begin{align*}
\rho_i(t+1) &= \rho_i(t) \quad \text{if } i\in\arg\max_{l\in N} f_l(t); \\
\rho_i(t+1) &= \rho_j(t) \quad \text{with}\ j\in N \ \text{such that}\  f_j(t) > f_i(t), \ \text{otherwise}.
\end{align*}
Note that this criterion fulfills \textit{NB} but violates \textit{SF}, as firms with the maximum average profit stick to their current rule rather than randomizing among all rules that yield maximal profit.\footnote{Of course, this could be fixed by adapting the \textit{Imitate-if-Better} criterion by imposing that a firm with maximal profit randomizes among all rules that also yield maximal profit.}  
\end{example} 

Our framework allows for heterogeneity among firms in rule revision. Thus, not all firms have to follow the same criterion, but rather could apply different criteria belonging to this class. For instance, some may follow the \textit{Imitate-the-Best-Max} criterion, others may adopt the \textit{Experimental} criterion, while still others may employ alternative criteria, as long as they respect the two basic principles.

Further, note that different criteria may have different informational requirements. While the \textit{Imitate-the-Best-Max} criterion requires firms to observe everybody else's average profit, other criteria may require only information on a subset of firms in the industry (e.g. \textit{Imitate-the-Best-Max with Sampling} criterion) or no information about average profits at all (e.g. \textit{Experimental} criterion). 

\subsection{Mistakes and Long-run equilibria}

The dynamics of action- and rule- revision gives rise to a time-homogeneous Markov process $\{\bm\omega(t)\}_{t\geq M}$ over a finite state space $\mathcal{S}$, where 
\[
\bm{\omega}(t) = \Bigg( \big( \bm{q}(r), \bm{\rho}(r), \bm{\tilde\tau}(r) \big) \Bigg)_{r = t - M + 1}^{t}.
\]
%, where each state is given by
%\[
%\bm{\omega}(t) = (\bm{q}(t),\ \bm{\rho}%(t),\ t\bm{1} - \bm{\tau}(t),\ \bm{f}(t)),
%\]
%and \(\bm{1} \in \mathbb{N}^n\) denotes the vector of ones.
%\footnote{Since our focus is on the system's long-run behavior, we initialize the system by assuming that, at the end of period 1, each firm $i$ randomly selects its future rule $\rho_i(2)$ from $\mathcal{R}$ and sets $\tau_i(2) = 1$.} 
This process is known as the \textit{unperturbed dynamics}, denoted by $P$. An \textit{absorbing set} of this process is a minimal subset $\Omega$ of the state space such that, once the dynamics enters $\Omega$, the probability of leaving it is zero. %As it turns out later, our process will only allow for singleton absorbing sets which are referred to as absorbing states.%
We denote the set of absorbing sets by $Z$ and use $z=|Z|$ to denote the number of absorbing sets. 

Note that once the unperturbed process reaches an absorbing set, it will stay there forever. In order to assess which kind of rules and actions will emerge in the long run, we consider a noisy version of the revision process where firms occasionally make mistakes. 
Specifically, there are two types of mistakes: \textit{action mistakes} and \textit{rule mistakes}. Action mistakes and rule mistakes are assumed to be independent of one another and independent across firms and over time. Action mistakes happen with fixed probability $\varepsilon \in (0,1)$. When such a mistake occurs an updating firm $i$ ignores the prescription of her rule and selects a quantity from $\Gamma$ at random, according to a pre-specified distribution with full support. Likewise with probability $\varepsilon^{\eta}$ a revising agent is subject to a mistake in rule choice, ignores her rule revision prescription and adopts a rule from $\mathcal{R}$ at random. The parameter $\eta$ captures the relative frequency of rule mistakes compared to action mistakes. We focus on $\eta \geq 1$, so that action mistakes occur more frequently than rule mistakes.\footnote{See also \cite{juang2002rule} and \cite{jackson2002formation} for contributions where mistakes in different components of agents' strategies arrive at different rates.} This assumption captures the intuitive idea that firms are less prone to mistakes when choosing rules for action revision than when applying these rules to choose specific actions.

\vspace{5pt}

The process with mistakes is referred to as the \textit{perturbed dynamics}, denoted by $P^{\varepsilon}$. For each $\varepsilon>0$, $P^{\varepsilon}$ is an irreducible and aperiodic Markov chain and has a unique invariant distribution $\mu(\varepsilon)$. It is a well-established result that $\mu^{*}=\lim_{\varepsilon\rightarrow 0}\mu{(\varepsilon)}$ exists and is an invariant distribution of the unperturbed dynamics $P$ \citep{freidlin1984random, kandori1993learning,young1993evolution}. The states in the support of $\mu^{*}$ are called \textit{long-run equilibria} (\textit{LRE}) and provide a prediction for the long-run behavior of the dynamic process for vanishing $\varepsilon$.

We identify LRE using an algorithm due to \cite{freidlin1984random}. We denote the transition cost in terms of the minimum number of mistakes required for the (direct) transition from one absorbing set $\Omega$ to another $\Omega'$ by $C(\Omega, \Omega')>0$.  An
$\Omega$-tree is a directed tree rooted into the absorbing set $\Omega$ where the nodes of this tree comprise all absorbing
sets. The cost of this tree is given by the sum of the transition cost  of each edge. Then an absorbing set $\Omega$ is LRE if and only if there exists
an $\Omega$-tree with minimal cost among all trees.\footnote{Strictly speaking, LRE is defined at the state level: all states in $\Omega$ are LRE if and only if the $\Omega$-tree has minimal cost among all trees; otherwise, none of them are. With a slight abuse of terminology, we describe such an absorbing set $\Omega$ as LRE.} The set of LRE is denoted by $Z'$.

\section{Results}
\label{section: results}
We are now able to present and discuss our results. To facilitate the exposition, we introduce the following notation. Let $\text{mon}(q, \rho)$ denote the set of states in which all firms employ rule $\rho$ and produce quantity $q$ over $M$ periods, and hence earn the same average profit $\pi(q, nq)$. Formally,
\[
\text{mon}(q, \rho) := \left\{  \big( (\bm{q}^{(r)}, \bm{\rho}^{(r)}, \bm{\tilde\tau}^{(r)}) \big)_{r=1}^M \in \mathcal{S} \;\middle|\; \bm{q}^{(r)} = (q, \dots, q),\; \bm{\rho}^{(r)} = (\rho, \dots, \rho), \text{ for all } r \right\}.
\]

Our first insight concerns the rules used by firms in absorbing sets of the unperturbed dynamics. 
\begin{lemma}
\label{lemma: 1}
All firms use the same behavioral rule in any absorbing set.    
\end{lemma}
Intuitively, for any state where firms use different behavioral rules, the \textit{Survival-of-the-Fittest} principle implies that there exists positive probability that all firms adopt the same rule. Given any such state, the \textit{No-Birth} principle ensures that no other rule can emerge. We remark that the reasoning does not depend on the strategic nature of the base game or the set of available behavioral rules. In this sense, Lemma~\ref{lemma: 1} can easily be adopted to a more general analysis of rule evolution. 

Our next result characterizes convergence of our process when all firms use $BR$.
\begin{lemma}\label{lemma: NE convergence}
From any state where all firms use $BR$ the process converges to $\text{mon}(q^N, BR)$ with probability one. 
\end{lemma}
We remark that if all firms choose their quantities according to $BR$, then by \textit{No-Birth} nobody will switch to $IM$. From there it follows that our process globally converges to  the set of states where all firms choose the Cournot Nash quantity $q^N$. In the proof we establish that the Cournot game under consideration is a finite game with strict strategic substitutes for which the best response dynamics with inertia in action revision converges to the unique Cournot Nash equilibrium. This result is an application of Theorem 2 in \cite{kukushkin2004best}.

Given the previous results, we are able to characterize the set of absorbing sets.

\begin{proposition}
\label{proposition: 1}
All firms use the same behavioral rule and produce the same quantity in
 any absorbing set. The set of absorbing sets is given by
\[
Z = \bigl\{\, \text{mon}(q, IM) \;\big|\, q \in \Gamma \bigr\}
   \bigcup
   \bigl\{\, \text{mon}(q^N, BR) \,\bigr\}.
\]
\end{proposition}

To appreciate the logic that underlies the proof of Proposition \ref{proposition: 1}, note that by Lemma \ref{lemma: 1}, there are two possible types of absorbing sets: (\romannumeral 1) sets of states in which all firms adopt \textit{IM}, and (\romannumeral 2)  sets of states in which all firms adopt \textit{BR}. Note that because of the \textit{No-Birth} principle in neither case can a new rule be introduced into the population. For case (i) note that, because of the \textit{Survival-of-the-Fittest} principle, in the absence of mistakes no new quantity emerges when all firms adopt $IM$ and choose the same quantity. 
For case (ii), Lemma \ref{lemma: NE convergence} establishes convergence to the Cournot Nash equilibrium.

We now proceed to present a series of lemmata that characterize the transitions between absorbing sets. For the sake of exposition we focus our discussion on transitions that are at the heart of our results and relegate the proof to Appendix \ref{Proof of Proposition 2}.

The first lemma is based on \cite{vega1997evolution}.
\begin{lemma}\label{lemma: Vega-redondo97}
Consider absorbing sets where all firms use $IM$. For any $q \neq q^W$ the transition costs are characterized by
$$C\left(\text{mon}(q, IM),\text{mon}(q^W, IM)\right)=1$$
and 
$$C\left(\text{mon}(q^{W}, IM),\text{mon}(q, IM)\right)>1.$$
\end{lemma}
The underlying idea is that the Walrasian quantity $q^{W}$ yields a strict relative-payoff advantage over any alternative quantity \citep{schaffer1988evolutionarily, vega1997evolution}. More formally, 
for all $ q \neq q^{W} $ and all $ 1 \leq m < n $,
\begin{equation}\label{eq: Walrasian quantity yields higher relative payoff}
p\big(m q^{W}+(n - m)q\big) q^{W} - c(q^{W}) > p\big(m q^{W}+(n - m)q\big) q - c(q).
\end{equation}
Thus, in an industry of imitators reaching the Walrasian equilibrium requires only a single action mistake, while leaving it requires more mistakes.

The next lemma generalizes the previous insight by arguing that in fact, in addition to the Walrasian quantity, other deviations away from symmetric output profiles may also improve the relative payoff advantage of a firm. To this end, denote the difference in profits between a single firm producing $q'$ and all other firms producing $q$ by
\begin{equation*}
    \Delta(q, q') = p(q' + (n-1)q)(q' - q) + c(q) - c(q').
\end{equation*}
Note that if $\Delta(q, q') \geq 0$, then a firm deviating from the output profile of $\text{mon}(q, IM)$ to $q'$ will enjoy a relative payoff advantage and may be imitated. Consequently denote by $D(q)=\{q'\in \mathbb{R}_{+} | \Delta(q, q') \geq 0\}$ the set of quantities that will be imitated with positive probability starting from a profile where everybody chooses $q$. 

The following lemma establishes important properties of the set $D(q)$ and discusses the implications for transition costs. 

\begin{lemma}\label{lemma: transitions by action mistake}
i) The set $D(q)$ is characterized as follows
$$D(q)=\begin{cases}
  [q,h(q)] & \text{if } q < q^W \\
  [\ell(q),q] & \text{if } q > q^W\\
  \{q^W\} & \text{if } q = q^W
\end{cases}$$
where $\ell(q)< q^W< h(q)$. ii) For any absorbing set $\text{mon}(q,IM)$ with $q \neq q^W$, we have
$$C\left(\text{mon}(q, IM),\text{mon}(q', IM)\right) =1$$
for all $q' \in \Gamma\cap(D(q)\setminus\{q\})$. 
\end{lemma}

\begin{figure}[h]
    \centering     \includegraphics[width=1\linewidth]{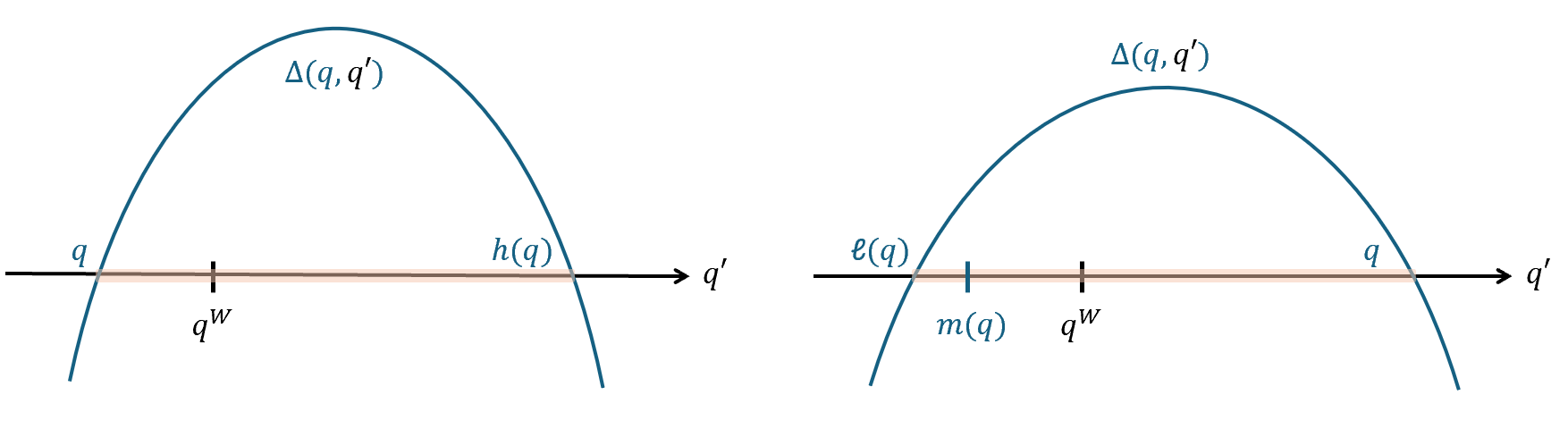}
    \caption{Illustration of the function $\Delta (q, q')$. The horizontal axis plots the variable $q'$. For any fixed $q<q^{W}$ ($q>q^{W}$), the curve in the left (right) panel depicts the value of $\Delta(q,q')$. The orange shade on the axis depicts the set $D(q)$. The definition of $m(q)$ shown in the right panel is provided in Appendix \ref{App: A}.} 
    \label{fig: delta}
    \vspace{5pt}
\end{figure}

In essence, when everybody chooses a quantity below the Walrasian quantity $q<q^W$, then certain quantity-increasing mistakes will be copied and when everybody chooses a quantity above the Walrasian quantity $q >q^W$, then certain quantity-decreasing mistakes will be followed. Figure \ref{fig: delta} illustrates this point by plotting the function $\Delta(q, q')$ and the set $D(q)$ for a 
quantity below the Walrasian quantity  and one above the Walrasian quantity. 

The next lemma characterizes transitions between absorbing sets where firms uniformly use $IM$ and the absorbing set where firms uniformly use $BR$. 
\begin{lemma}\label{lemma: transitions by rule mistake}
Consider the absorbing sets $\text{mon}(q^N, BR)$, $\text{mon}(q^N, IM)$, and $\text{mon}(q^W, IM)$. Then 
$$C\left(\text{mon}(q^N, BR),\text{mon}(q^N, IM)\right) = C\left(\text{mon}(q^{W}, IM),\text{mon}(q^N, BR)\right)= \eta \geq 1.$$
\end{lemma}
Thus, both of these transitions only require one rule mistake (at the cost of $\eta$). To see this, start with the transition from $\text{mon}(q^N, BR)$ to  $\text{mon}(q^N,IM)$. Assume 
that firm $i$ makes a rule mistake and switches to $IM$. Since there is no action mistake, firm $i$ continues to play $q^N$, yielding the same profit as all other firms. Because of the \textit{Survival-of-the-Fittest} principle there is a positive probability that all other firms come to adopt $IM$ and the process reaches $\text{mon}(q^N,IM)$. For the transition from $\text{mon}(q^{W}, IM)$ to $\text{mon}(q^N,BR)$  there are actually two alternative paths which both only require one mistake in rule choice:
For the first path consider $\text{mon}(q^{W}, IM)$ and assume (due to inertia in rule choice)  firms have not revised their rules for the past $x$ periods ($1\leq x<M$). From there consider a rule mistake by firm $i$ which immediately chooses the best response quantity which we denote by $\tilde q = b_i\left((n-1)q^W\right)$.\footnote{Without loss of generality, take $\tilde{q}\in \Gamma$. More details are provided in Appendix \ref{App: A}.} 
Now observe that the average past performance of $i$'s rule is $f_i=\pi\left(\tilde q,(n-1)q ^W+\tilde q\right)>\pi^W$. In contrast, for all other firms the average past performance of their rule is given by $f_j=\frac{x\pi^W+\pi(q^W,(n-1)q^W+\tilde q)}{x+1}$. For sufficiently large $x$, we have $f_i>f_j$ and so other firms will adopt the $BR$ rule with positive probability. Regarding the second path, consider $\text{mon}(q^{W}, IM)$  and assume firm $i$ makes a mistake and switches its rule to $BR$ but (due to inertia in action revision) does not adjust its quantity yet. Now both $IM$ and $BR$ yield the same past average performance and due to \textit{Survival-of-the-Fittest} with positive probability all other firms switch to $BR$.
Note that while the first path utilizes inertia in rule revision, the second one is based on inertia in quantity revision.\footnote{Thus our results would also hold under alternative revision processes that  model inertia in a more general way than in the present exposition. Moreover, this indicates that our results do not hinge on intertemporal comparisons of payoffs, which in themselves may introduce a force that pushes away from the Walrasian equilibrium (see \citealt{alos2004cournot}).}

With these insights at hand, we are able to state our main result.

\begin{proposition}
\label{proposition: 2}
The set of long-run equilibria is characterized as follows. 
\begin{enumerate}[label=(\roman*), itemsep=-10pt]
    \item If $\eta > 1$, then
    \[
   Z'= \text{mon}(q^{N}, BR) \bigcup  \text{mon}(q^{W}, IM),
    \]
    \item If $\eta = 1$,  there exist thresholds $\underline{q}$ and $\overline{q}$ satisfying $\underline{q} \leq q^{N} < q^{W} \leq \overline{q}$, such that 
    \[
    Z' = \text{mon}(q^{N}, BR)\bigcup \big(\bigcup_{q \in \Gamma \cap [\underline{q}, \overline{q}]} \text{mon}(q, IM)\big).
    \]
\end{enumerate}
\end{proposition}

When rule mistakes occur less frequently than action mistakes, $\eta >1$, the logic is as follows.  By lemmata \ref{lemma: Vega-redondo97} and \ref{lemma: transitions by rule mistake} we can exhibit $\text{mon}(q^{N}, BR)$ and  $ \text{mon} (q^{W}, IM)$-trees of cost $z+\eta-2$ where  $z$ is the number of absorbing sets. Because of the {\it No-Birth} principle, we can only move between absorbing sets where firms choose different rules with a rule mistake (at cost $\eta>1$). It follows that these two trees are of minimum cost. Moreover, any other tree has to include a branch going out of $\text{mon}(q^{N}, BR)$ at cost $\eta$ and a branch going out of  $\text{mon} (q^{W}, IM)$ at cost strictly larger than $1$. It follows that $\text{mon}(q^{N}, BR)$   $\bigcup  \text{mon} (q^{W}, IM)$ is the unique LRE. 

When rule mistakes occur at the same frequency as action mistakes, $\eta =1$, it is still true that there exist $\text{mon}(q^{N}, BR)$ and  $\text{mon} (q^{W}, IM)$-trees of minimum cost $z-1$. However, there now exist several other  $ \text{mon} (q, IM)$-trees with minimal cost. For instance, consider $\tilde q=h(q^N)>q^{W}$. Because of lemma \ref{lemma: transitions by action mistake} we can move from $\text{mon}(q^{N}, IM)$ to  $\text{mon}(\tilde q,IM)$ with one mistake.
It follows there exists a $\text{mon}(\tilde q,IM)$-tree the  cost of which is also equal to $z-1$.\footnote{
In this tree all other absorbing states are connected to $\text{mon}(q^{W}, IM)$ at the cost of one mistake each and $\text{mon}(q^{W}, IM)$ is connected to $\text{mon}(q^{N}, BR)$ at the cost of one mistake.} Using the same and related ideas, we can use the characterization provided in Lemma \ref{lemma: transitions by action mistake} to identify the entire set of long run equilibria.

\begin{rmk}
The exact values of $\underline{q}$ and $\bar{q}$ depend on the number of firms in the industry. Specifically, if $n=2$ (a duopoly), then $\underline{q}=q^{N}$ and $\bar{q}=h(q^{N})$; if $n\geq3$, and the grid of feasible output $\Gamma$ is sufficiently fine, then $\underline{q}=0$ and $\bar{q}=h(0)$. The detailed proof is relegated to Appendix \ref{Proof of the Remark}. 
\end{rmk}

To facilitate understanding of our results, we present the following example. 

\begin{example}
\label{example: a Cournot oligopoly}
Consider $n=4$ firms, an inverse demand function of $p(Q)=\max{\{90-Q, 0\}}$, and a cost function of $c(q)=\frac{1}{2}q^2$. Let the grid of feasible output be $\Gamma = \{0, 1, \dots, 90\}$. The Nash quantity is $q^{N} = 15$, while the Walrasian quantity is $q^{W} = 18$. The collusive quantity, defined as $q^{C} := \displaystyle \arg\max\limits_{q \in \mathbb{R}_+} \pi(q,nq)$, is $q^{C} = 10$. If $\eta>1$, then $Z'= \text{mon}(18, IM)\bigcup\text{mon}(15, BR)$ and if $\eta = 1$, then $Z'=\big(\bigcup_{q \in \{0, 1, \dots, 60\}} \text{mon}(q, IM)\big)\bigcup\text{mon}(15, BR)$. 
\end{example}  
    
Several intriguing points stand out. First, $\text{mon}(q^{N}, BR)$ is always LRE even though imitators earn higher relative profits than best responders in mixed profiles \citep{schipper2009imitators}. Since all firms respect the \textit{Survival-of-the-Fittest} principle when revising their behavioral rules, the rule that yields higher profit tends to be attractive. Nevertheless, our results show that states in which firms uniformly adopt the $BR$ rule can still emerge in the long run when firms may revise their rules. Second, when rule mistakes occur at the same frequency as action mistakes ($\eta = 1$), our findings reveal a sharp contrast to the unique Walrasian outcome predicted in \cite{vega1997evolution} under uniform adoption of the $IM$ rule. Specifically, the introduction of rule revision leads to a broad spectrum of long-run output. Firms' output can persist strictly below $q^{N}$ when the industry comprises at least three firms. Notably, the output level may fall to the \textit{collusive} quantity, in which case the industry achieves maximal efficiency.

\section{Conclusion}
\label{section: conclusion}

This paper highlights the implications of endogenous rule revision in the Cournot oligopoly. Although our analysis has been developed in the context of Cournot oligopolies, the mechanisms underlying our results are not confined to this particular setting. Rather, they rely on the general property of quasi-submodularity, which characterizes a broad class of aggregative games \citep{corchon1994comparative}. Prominent examples within this category include Cournot oligopoly, the tragedy of the commons, rent-seeking games, and common-pool resource games \citep{, schipper2004submodularity, alos2005evolutionary}. In Cournot terms, quasi-submodularity implies that a firm's payoff advantage of a higher output over a lower one preserves as the aggregate quantity decreases, which can be viewed as a form of strategic substitute. Crucially, extending the results to aggregative games with strict quasi-submodularity does not alter the economic implications of our findings. Appendix \ref{App: B} provides an extensive discussion.

One promising avenue for future research is to examine the robustness of our results by incorporating the costs of behavioral rules \citep{conlisk1980costly,droste2002endogenous,oprea2020makes}. Best response, being more computationally demanding, may entail higher implementation costs, whereas imitation learning, which relies on information about both actions and payoffs, may involve greater information costs. Another important direction is to investigate alternative principles of rule revision, particularly those of an uncoupled or reinforcement nature \citep{fatas2024social}. Such principles rely solely on personal information, in contrast to our principles grounded in imitative heuristics. Studying such models may provide valuable insights into the evolution of rules in environments where the rules adopted by others are unobservable or difficult to infer.

\newpage

\begin{comment}

\textcolor{red}{[ZBY: Some points for further studies: (1) Other imitation dynamics, such as imitate-if-better and pairwise comparison. (2) Network game (players can only observe the actions and payoffs of some co-players). In a network Cournot game, IM also converges to $q^W$ (Shi, Zhang, 2019, Econ Lett). What happens for BR? (3) Other behavioral rules, such as unconditional behavioral rules (i.e., unconditionally choose an action). Do adaptive behavioral rules such as IM and BR always dominate unconditional rules?} \textcolor{brown}{(1)Imitate-if-better, Pairwise comparison or PIR will make a difference, but under these rules it is hard to derive LRE. (2) I think it is promising to extend the model to network. Maybe we can consider multiple markets, and introduce information networks on the observation of ``fitness''. (3) Combinations of different games and different plausible rules remain further investigation.}

\textcolor{red}{[ZBY: It seems that we cannot simply conclude that rule mistake promotes/stabilizes cooperation. When $\eta=1$, the LRE set also includes $q<q^C$ and $q>q^W$. Thus, it is not clear whether the expected quantity (at the stationary distribution) is lower than $q^C$.\textcolor{brown}{Please see the Remark. Besides, I used the word ``can''. } Besides, how do we understand this result? Can we say that mistakes in rule revision make the long-run outcome unpredictable.]}
\end{comment}

\newpage
\appendix
\section{Appendix: Main Proofs}\label{App: A}
\renewcommand{\thefigure}{A.\arabic{figure}}
\setcounter{figure}{0}
\renewcommand{\theequation}{A.\arabic{equation}}
\setcounter{equation}{0}

\subsection{}
\label{Proof of Proposition 1}
\begin{proof}[\textbf{Proof of Lemma \ref{lemma: 1}}]
\label{proof of lemma 1}
Consider any state in the state space, $\bm\omega \in \mathcal{S}$. After the base game is played, suppose that all firms receive opportunities to revise their behavioral rules. Let $i^*$ be a firm with the highest average profit in the industry, i.e., $i^* \in \arg\max_{l \in N} f_l$. Note that, under the \textit{Survival-of-the-Fittest} principle, each firm with positive probability switches to the rule employed by a firm with average profit no less than its own. In particular, each firm may switch to the rule currently employed by firm $i^{*}$. In that case, all firms adopt the same behavioral rule in the next period. Further, by the principle of \textit{No-Birth}, no new rule emerges thereafter. Therefore we conclude that from any initial state, there exists an unperturbed revision path leading to a state in which all firms adopt the same behavioral rule.
\end{proof}

\vspace{5 pt}

\begin{proof}[\textbf{Proof of Lemma \ref{lemma: NE convergence}}]
\label{proof of NE convergence}

We first show that a strictly decreasing inverse demand function $p(\cdot)$ implies that the Cournot oligopoly with a finite strategy set satisfies the condition of \textit{strict strategic substitutes} in the sense of \cite{kukushkin2004best}.\footnote{This condition is encompassed by the definition of \textit{strict quasi-submodularity}. See \cite{schipper2009imitators}. } Specifically, for  any two feasible output levels $q^{(1)}$ and $q^{(2)}$ with  $q^{(2)}>q^{(1)}$ and  associated  total industry output levels $Q^{(2)}>Q^{(1)}$, 
 a Cournot oligopoly satisfies the strict strategic substitutes condition if
\[
\pi(q^{(2)},Q^{(2)})\geq\pi(q^{(1)},Q^{(2)}) \quad\text{implies}\quad \pi(q^{(2)},Q^{(1)})>\pi(q^{(1)},Q^{(1)}). 
\]
To see this, note that the strictly decreasing $p(\cdot)$ implies 
$
p(Q^{(1)})>p(Q^{(2)})$. Hence, for any $q^{(2)}>q^{(1)}$, we have 
\begin{equation}
\label{eq: 1}
p(Q^{(1)})(q^{(2)}-q^{(1)})>p(Q^{(2)})(q^{(2)}-q^{(1)}).
\end{equation}
Now, suppose that $ \pi(q^{(2)},Q^{(2)})\geq\pi(q^{(1)},Q^{(2)})$. By the definition of profit, it follows that
\begin{equation}
\label{eq: 2}
p(Q^{(2)})(q^{(2)}-q^{(1)})\geq c(q^{(2)})-c(q^{(1)}).    
\end{equation}
Combining \ref{eq: 1} and \ref{eq: 2}, we obtain
$
p(Q^{(1)})(q^{(2)}-q^{(1)})> c(q^{(2)})-c(q^{(1)}),
$
which implies that  $\pi(q^{(2)},Q^{(1)})>\pi(q^{(1)},Q^{(1)})$.

Now, consider any state $\omega \in \mathcal{S}$ in which all firms employ $BR$ and assume that (due to inertia) in each period, only one firm adjusts its output. Then, based on Theorem 2 in \cite{kukushkin2004best}, after a finite number of periods, the sequence of output profiles converges to equilibrium. Moreover, under our assumptions, the Cournot Nash equilibrium is unique. Hence, all firms eventually produce the Nash quantity $q^{N}$, and no firm has an incentive to deviate thereafter. Therefore, we conclude that from any state where all firms use \textit{BR} the unperturbed dynamics $P$ converges globally to $\text{mon}(q^{N},BR)$.
\end{proof}

\vspace{5 pt}

\begin{proof}[\bf{\textit{Proof of Proposition \ref{proposition: 1}}}]
\label{proof of proposition 1}
First, it is straightforward that $\text{mon}(q, IM)$ for any $q \in \Gamma$, as well as $\text{mon}(q^N, BR)$, are absorbing sets. It remains to show that the unperturbed dynamics $P$ converges to one of these sets.

According to Lemma \ref{lemma: 1}, starting from any state in the state space, after finite periods, the dynamics will reach a state in which all firms adopt either $IM$ or $BR$. On one hand, if all firms adopt $IM$, then with positive probability, they switch to the output that yields the highest profit in the current period. As a result, all firms produce the same quantity from then on and eventually the dynamics reaches some $\text{mon}(q,IM)$, $q\in \Gamma$. On the other hand, if all firms adopt $BR$, then as established in Lemma \ref{lemma: NE convergence}, all firms eventually produce the Nash quantity and the dynamics reaches $\text{mon}(q^{N},BR)$. 
\end{proof}

\subsection{}
\label{Proof of Proposition 2}

\begin{proof}[\textbf{Proof of Lemma \ref{lemma: transitions by action mistake}}]
\label{proof of transitions by action mistake}
To characterize the set $D(q):=\{q'\in \mathbb{R}_{+}\mid\Delta(q,q')\geq0 \}$, we first illustrate the following properties of the function $\Delta(q,q') := p(q' + (n-1)q)(q' - q) + c(q) - c(q')$. Specifically, \textit{(i)} for $\forall q$, $\Delta(q, q)=0$;
 \textit{(ii)} $\forall q\in\mathbb{R}_{+}\setminus\{q^{W}\}$, $\Delta(q, q^{W})>0$;  
\textit{(iii)}  $\forall q \in [0, q^{W})$, $\Delta(q, q')$ is strictly concave in $q'\in [q,Q_{max}-(n-1)q]$. 

Note that \textit{(i)} follows directly from the definition of $\Delta$, and \textit{(ii)} is an immediate implication of Lemma \ref{lemma: Vega-redondo97}. Regarding \textit{(iii)}, note that 
\begin{equation}\label{eq: second-derivative-of-delta}
\frac{\partial^2 \Delta(q, q')}{\partial q'^2} = p''\left(q' + (n-1)q\right)(q' - q) + 2p'\left(q' + (n-1)q\right) - c''(q')
.
\end{equation}
Because  $p^{\prime\prime}\leq0$, $p^{\prime}<0$, and $c^{\prime\prime}\geq0$, \ref{eq: second-derivative-of-delta} implies \textit{(iii)}. \\

We  now show that the set $D(q)$ is characterized as 
$$D(q)=\begin{cases}
  [q,h(q)] & \text{if } q < q^W \\
  [\ell(q),q] & \text{if } q > q^W\\
  \{q^W\} & \text{if } q = q^W
\end{cases}$$
where $\ell(q)< q^W< h(q)$. We proceed by analyzing the three cases in turn.\\

\noindent \textbf{Case 1}\quad For any fixed $q \in [0,q^{W})$ properties \textit{(i)}-\textit{(iii)} imply that the equation $\Delta(q, q') = 0$ has exactly two solutions. One is the trivial solution $q' = q$; the other, denoted by $q' = h(q)$, satisfies $h(q) > q^{W}$ and $h(q)<Q_{max}-(n-1)q$. Hence, the function $h: [0, q^{W}) \to (q^{W}, Q_{max})$ is well defined. As illustrated by the left panel of Figure \ref{fig: delta}, $\Delta(q,q')\geq0$ if and only if $q'\in[q,h(q)]$. Therefore, we conclude that $D(q)=[q,h(q)]$ for $q<q^{W}$, where $h(q)>q^{W}$.\\

\noindent \textbf{Case 2}\quad 
For any fixed $q >q^{W}$,  properties \textit{(i)}-\textit{(iii)} imply that $\Delta(q,q')<0$ for all $q'>q$. Hence, $D(q)\subseteq[0,q]$. To further characterize $D(q)$  we denote by  $q'=m(q)\geq0$  the unique solution of $p(q'+(n-1)q)-c'(q)=0$.\footnote{Without loss of generality, we assume that $p((n-1)q)-c'(0)>0$. If $p((n-1)q)-c'(0)\leq0$, it is clear that $D(q)=[0,q]$.} Since $q>q^{W}$, it is obvious that $m(q)<q^{W}$. See the right panel of Figure \ref{fig: delta} for an illustration.

Note that for all $q'\in[m(q),q]$, 
\begin{equation}\label{eq: inequality related to m(q)}
    \Delta(q,q')\geq c'(q')(q'-q)+c(q)-c(q')= (c'(\xi)-c'(q'))(q-q')\geq 0,
\end{equation}
where $q'\leq\xi\leq q$. Thus, $[m(q),q]\subseteq D(q)$, for all $q >q^{W}$. Further, note that when $q'\in[0, m(q)]$, 
\begin{equation}\label{eq: first-derivative-of-delta}
    \frac{\partial \Delta(q,q')}{\partial q'}=p'(q'+(n-1)q)(q'-q)+p(q'+(n-1)q)-c'(q')>0.
\end{equation}
Thus, $\Delta(q,q')$ strictly increases in $[0,m(q)]$, and hence there exists unique $\ell(q)\in[0,m(q)]$ such that $\Delta(q,q')\geq0$ if and only if $q'\in [\ell(q), q]$. Specifically, if $\Delta(q,0)\leq0$, $\ell(q)\in[0,m(q)]$ satisfies $\Delta(q, \ell(q))=0$; otherwise, if $\Delta(q,0)>0$, then $\ell(q)=0$. It follows that for any $q>q^{W}$, $D(q)=[\ell(q),q]$, where $\ell(q)<q^{W}$.\footnote{Unlike in case 1, in case 2, $\Delta(q, q')$ is not necessarily strictly concave in $q' \in [0, q]$, and therefore the proof cannot proceed in the same way as in case 1.} \\
\noindent\textbf{Case 3}\quad  Lemma \ref{lemma: Vega-redondo97} implies $D(q^{W})=\{q^{W}\}$. 

\vspace{10pt}
Finally,  Lemma \ref{lemma: transitions by action mistake} ii) follows directly from Lemma \ref{lemma: transitions by action mistake} i).
\end{proof}

\vspace{5pt}
\begin{proof}[\textbf{Proof of Lemma \ref{lemma: transitions by rule mistake}}]
\label{proof of transitions by rule mistake}
We first show that a single rule mistake can induce a transition from $\text{mon}(q^{N}, BR)$ to $\text{mon}(q^{N}, IM)$. The reasoning proceeds as follows. Consider any state in $\text{mon}(q^{N}, BR)$. From there suppose that firm $i$ revises its rule and mistakenly adopts $IM$. Accordingly, firm $i$ updates its output by imitation and hence still produces the Nash quantity $q^{N}$. Since all firms continue to choose the Nash quantity, they share the same average profit $\pi^{N}$, regardless of the length of the evaluation window for each firm. As a result, $BR$ and $IM$ yield the same average performance. By the \textit{Survival-of-the-Fittest} principle, all other firms may subsequently adopt $IM$ as well. Regarding the output level, either by imitation or inertia, all firms continue to produce $q^{N}$. Therefore, we conclude that $C(\text{mon}(q^{N}, BR), \text{mon}(q^{N}, IM)) = \eta$.

Next, we show that one rule mistake can lead to the transition from $\text{mon}(q^{W},IM)$ to $\text{mon}(q^{N},BR)$. There are actually two possible revision paths. For the first path, starting from any state in $\text{mon}(q^{W}, IM)$, consider that the dynamics reaches a state in which no firm has revised its rule for the past $x$ periods ($1\leq x<M$) due to inertia in rule choice. From there consider a rule mistake by firm $i$ which immediately chooses a best response quantity, denoted by $\tilde{q}\in \Gamma$. Then two cases arise. On one hand, if $q^{W}$ is not a best response to the Walrasian equilibrium, the average past performance of $i$'s current rule (i.e., $BR$) exceeds $i$'s previous profit under the Walrasian outcome. That is, $f_i=\pi\left(\tilde q,(n-1)q ^W+\tilde q\right)>\pi^{W}$. Meanwhile, for all other firms, the average past performance of their rule (i.e., $IM$) is given by $f_j=\frac{x\pi^W+\pi(q^W,(n-1)q^W+\tilde q)}{x+1}$. For sufficiently large $x$, we have $f_i>f_j$ and so other firms will adopt the $BR$ rule with positive probability. On the other hand, if $q^{W}$ is a best response to the Walrasian equilibrium (which may occur when the grid $\Gamma$ is sufficiently coarse), then firm $i$ may still choose the Walrasian quantity. Since all firms continue to choose the Walrasian quantity, they share the same average profit $\pi^{W}$ for any $x$. As a result, $IM$ and $BR$ yield the same average performance and all firms may subsequently adopt $BR$.

Regarding the second path, consider $\text{mon}(q^{W},IM)$ and assumes that a firm mistakenly switches to $BR$ but (due to inertia in action choice) does not adjust its quantity yet. Note that $IM$ and $BR$ yield the same average performance and because of the  \textit{Survival-of-the-Fittest} principle with positive probability all other firms switch to $BR$. In light of the two paths mentioned above, by Lemma 2, the dynamics finally converges to $\text{mon}(q^{N}, BR)$. We further conclude that $C(\text{mon}(q^{W}, IM), \text{mon}(q^{N}, BR)) = \eta$.

\end{proof}

\vspace{5 pt}

\begin{proof}[\textbf{Proof of Proposition \ref{proposition: 2}}]
\label{proof of proposition 2}
With Lemmata \ref{lemma: transitions by action mistake} and \ref{lemma: transitions by rule mistake} established, we now characterize the long-run equilibria.
\par \noindent \textbf{Case 1}
We first consider the case where $\eta>1$. Note that the transition cost of any directed edge starting from or ending at $\text{mon}(q^{N}, BR)$ must be at least $\eta\geq1$. Therefore, for any absorbing set $\Omega$ and any $\Omega$-tree, the cost of the tree must be no less than $z-2+\eta$, where $z$ denotes the number of absorbing sets.

Next, we show that there exists a $\text{mon}(q^{W}, IM)$-tree and a $\text{mon}(q^{N}, BR)$-tree, each with cost equal to $z-2+\eta$. Specifically, the corresponding trees are illustrated below:

\vspace{5 pt}

\begin{center}
\begin{tikzpicture}[>=stealth, node distance=1cm]
    \node (A) at (-4,2) {$\text{mon}(q^{W}, IM)$};
    \node (B) at (0,2) {$\text{mon}(q^{N}, IM)$};
    \node (C) at (4,2) {$\text{mon}(q^{N}, BR)$};
    \node (D) at (-4,0) {$\text{mon}(\tilde{q}, IM)$};

    \draw[->] (B) -- node[above] {1} (A);
    \draw[->] (C) -- node[above] {$\eta$} (B);
    \draw[->] (D) -- node[left] {1} (A);    
\end{tikzpicture}\\[1cm]
\begin{tikzpicture}[>=stealth, node distance=2cm]
    \node (A) at (-4,2) {$\text{mon}(q^{N}, BR)$};
    \node (B) at (0,2) {$\text{mon}(q^{W}, IM)$};
    \node (C) at (4,2) {$\text{mon}(q^{N}, IM)$};
    \node (D) at (0,0) {$\text{mon}(\tilde{q}, IM)$};

    \draw[->] (B) -- node[above] {$\eta$} (A);
    \draw[->] (C) -- node[above] {1} (B);
    \draw[->] (D) -- node[left] {1} (B);  
\end{tikzpicture}
\end{center}
\vspace{-0.5em}
Here, $\tilde{q}$ denotes any quantity that belongs to $\Gamma\setminus\{q^{W},q^{N}\}$. As an application of the \cite{freidlin1984random} algorithm, it follows that both $\text{mon}(q^{N}, BR)$ and $\text{mon}(q^{W}, IM)$ are LRE. 

Further, we show that the cost of any $\text{mon}(\tilde{q}, IM)$-tree with $\tilde{q} \neq q^{W}$ is strictly greater than $z - 2 + \eta$. The reasoning proceeds as follows.  Since $\text{mon}(\tilde{q}, IM)$ is the terminal node of any $(\tilde{q}, IM)$-tree, the tree must include an edge starting from $\text{mon}(q^{W}, IM)$ and another starting from $\text{mon}(q^{N}, BR)$. The cost of the former is strictly greater than 1, and the cost of the latter is no less than $\eta$. Therefore, the cost of any $\text{mon}(\tilde{q}, IM)$-tree with $\tilde{q} \neq q^{W}$ is strictly greater than $z - 2 + \eta$. As an application of the algorithm in \cite{freidlin1984random}, it follows that $\text{mon}(\tilde{q}, IM)$ is not LRE, for all $\tilde{q}\neq q^{W}$. We further conclude that when $\eta>1$, $Z' =  \text{mon}(q^{N}, BR)\bigcup \text{mon}(q^{W}, IM)$. 
\vspace{10 pt}
\par \noindent \textbf{Case 2}\quad Consider the case where $\eta=1$. Then, for any absorbing set $\Omega$ and any $\Omega$-tree, the cost of the tree must be at least $z-1$. We further demonstrate that there exists a $\text{mon}(q^{W}, IM)$-tree, a $ \text{mon}(q^{N}, BR)$-tree and a $\text{mon}(q^{N}, IM)$-tree, each with cost equal to $z-1$. Specifically, the corresponding trees are shown below:

\vspace{5 pt}

\begin{center}
\begin{tikzpicture}[>=stealth, node distance=1cm]
    \node (A) at (-4,2) {$\text{mon}(q^{W}, IM)$};
    \node (B) at (0,2) {$\text{mon}(q^{N}, IM)$};
    \node (C) at (4,2) {$\text{mon}(q^{N}, BR)$};
    \node (D) at (-4,0) {$\text{mon}(\tilde{q}, IM)$};

    \draw[->] (B) -- node[above] {1} (A);
    \draw[->] (C) -- node[above] {1} (B);
    \draw[->] (D) -- node[left] {1} (A); 
\end{tikzpicture}
\\[1cm]
\begin{tikzpicture}[>=stealth, node distance=2cm]
    \node (A) at (-4,2) {$\text{mon}(q^{N}, BR)$};
    \node (B) at (0,2) {$\text{mon}(q^{W}, IM)$};
    \node (C) at (4,2) {$\text{mon}(q^{N}, IM)$};
    \node (D) at (0,0) {$\text{mon}(\tilde{q}, IM)$};

    \draw[->] (B) -- node[above] {1} (A);
    \draw[->] (C) -- node[above] {1} (B);
    \draw[->] (D) -- node[left] {1} (B);   
\end{tikzpicture}
\\[1cm]
\begin{tikzpicture}[>=stealth, node distance=2cm]
    \node (A) at (-4,2) {$\text{mon}(q^{N}, IM)$};
    \node (B) at (0,2) {$\text{mon}(q^{N}, BR)$};
    \node (C) at (4,2) {$\text{mon}(q^W, IM)$};
    \node (D) at (4,0) {$\text{mon}(\tilde{q}, IM)$};

    \draw[->] (B) -- node[above] {1} (A);
    \draw[->] (C) -- node[above] {1} (B);
    \draw[->] (D) -- node[left] {1} (C);   
\end{tikzpicture}
\end{center}
Here, $\tilde{q}$ denotes any quantity that belongs to $\Gamma\setminus\{q^{W},q^{N}\}$. From the \cite{freidlin1984random} algorithm, it follows that $\text{mon}(q^{W}, IM)$, $ \text{mon}(q^{N}, BR)$, and $\text{mon}(q^{N}, IM)$ are LRE. 
\vspace{-0.5em}

Further, by Lemma \ref{lemma: transitions by action mistake}, the transition cost is $C(\text{mon}(q^N, IM), \text{mon}(q', IM))=1$ for any $q'\in\Gamma\cap(q^{N}, h(q^{N})]$. Hence, there exists a $\text{mon}(q', IM)$-tree with cost $x-1$. Specifically, 
\begin{center}
\begin{tikzpicture}[>=stealth, node distance=2cm]
    \node (A) at (-4,2) {$\text{mon}(q^{N}, IM)$};
    \node (B) at (0,2) {$\text{mon}(q^{N}, BR)$};
    \node (C) at (4,2) {$\text{mon}(q^W, IM)$};
    \node (D) at (4,0) {$\text{mon}(\tilde{q}, IM)$};
    \node (E) at (-4,0) {$\text{mon}({q'}, IM)$};
    \draw[->] (A) -- node[left] {1} (E);    
    \draw[->] (B) -- node[above] {1} (A);
    \draw[->] (C) -- node[above] {1} (B);
    \draw[->] (D) -- node[left] {1} (C);   
\end{tikzpicture}
\end{center}
\vspace{-0.5em}
where $\tilde{q}$ denotes any quantity in $\Gamma \setminus \{q', q^{W}, q^{N}\}$. The \cite{freidlin1984random} algorithm, implies that $\text{mon}(q, IM)$ is LRE for all $q\in\Gamma\cap[q^{N}, h(q^{N})]$, where $h(q^{N})>q^{W}$. We now define
$$
\underline{q} := \min \left\{ q \in \Gamma \mid \text{mon}(q, IM)\ \text{is LRE} \right\}, \quad
\bar{q} := \max \left\{ q \in \Gamma \mid \text{mon}(q, IM)\ \text{is LRE} \right\}.
$$ Based on the previous analysis, we obtain that $\underline{q}\leq q^{N}$ and $\bar{q}\geq q^{W}$. Further, by Lemma \ref{lemma: transitions by action mistake}, $\Delta(\underline{q},q')\geq 0$ for all $q'\in(\underline{q},q^{N})$, if such $q'$ exists. Hence, a single action mistake can lead to the transition from $\text{mon}(\underline{q},IM)$ to $\text{mon}(q',IM)$, for all $q'\in\Gamma\cap(\underline{q},q^{N})$. Besides, a single action mistake can also lead to the transition from $\text{mon}(q',IM)$ to $\text{mon}(q^{W},IM)$. As an application of Theorem 3 in \cite{ellison2000basins}, which is referred to as the \textit{tree surgery} argument, we conclude that $\text{mon}(q',IM)$ is LRE for all $q'\in\Gamma\cap(\underline{q},q^{N})$. 

Further, Lemma \ref{lemma: transitions by action mistake} implies that $\Delta(\bar{q},q')\geq 0$ for all $q'\in(q^{W},\bar{q})$, if such $q'$ exists. Following the same reasoning as above, we conclude that $\text{mon}(q',IM)$ is LRE for all $q'\in\Gamma \cap (q^{W},\bar{q})$. The proof is thus complete.
\end{proof}

\subsection{}
\label{Proof of the Remark}
We first point out additional properties of $\Delta(\cdot,\cdot)$ that are important for understanding our results, continuing the numbering in \ref{proof of transitions by action mistake}. Specifically, \textit{(iv)} when $n=2$, $\Delta(q,q')=\Delta(q',q)$, for $\forall q, q'$; \textit{(v)} when $n>2$, $\Delta(q,q')+\Delta(q',q)>0$, for $\forall q\neq q'$; \textit{(vi)} the functions $h(\cdot)$ and $\ell(\cdot)$ defined in Lemma~\ref{lemma: transitions by action mistake} are continuous and strictly decreasing.  These properties follow directly from the definitions and the Implicit Function Theorem.

\vspace{0.5em}
\noindent \textbf{Case 1}\quad Consider the case of $n = 2$ (duopoly). We first show that a single action mistake cannot induce a transition from $\text{mon}(q,\text{IM})$ to $\text{mon}(q',\text{IM})$, for any $q\in \Gamma\cap[q^{N},h(q^{N})]$ and $q'\notin \Gamma\cap[q^{N},h(q^{N})]$. To see this, note that by \textit{(iv)}, we have 
\begin{equation}
\label{eq: n = 2}
\Delta(q^{N},h(q^{N}))=\Delta(h(q^{N}),q^{N})=0. 
\end{equation}
Moreover, by \textit{(vi)}, $h(q) < h(q^N)$ for any $q \in (q^N, q^W)$, implying that $D(q)\subset[q^{N},h(q^{N})]$. Similarly, \textit{(vi)} also implies that $\ell(q) > \ell(h(q^N))$ for any $q \in (q^W, h(q^N))$. Since Formula \eqref{eq: n = 2} implies $\ell(h(q^N)) = q^N$, it follows that $\ell(q) > q^N$, so again $D(q)\subset[q^{N},h(q^{N})]$. 

Next, we demonstrate that a single rule mistake cannot lead to a transition from $\text{mon}(q,\text{IM})$ to $\text{mon}(q',\text{IM})$, for any $q\in \Gamma\cap[q^{N},h(q^{N})]$ and $q'\notin \Gamma\cap[q^{N},h(q^{N})]$. The reasoning proceeds as follows. Starting from any state in $\text{mon}(q,\text{IM})$, suppose a single rule mistake occurs. Thereafter, along any unperturbed revision path, firms' output must be either $q$ or some iterated best response to $q$, denoted by $\tilde{q}\in \Gamma$.  Since the best response correspondence is a contraction in the duopoly setting \citep[][Chapter 4]{vives1999oligopoly}, it follows that $\tilde{q} \leq q$.

Moreover, without loss of generality, assume that at least one firm follows $IM$ in each period for any unperturbed revision path. Consider any period in which firm 1 produces $q_1$ and firm 2 produces $q_2$. If $q\geq q_{1}\geq q_{2}\geq q^{N}$, then one firm will produce $q_1$ by imitation, or produce either $q_{1}$ or $q_{2}$ due to inertia. If instead $q\geq q_{1}\geq q^{N}>q_{2}$, then if firm $1$ follows $IM$, it will again produce $q_1$ via imitation or inertia; if firm $1$ follows $BR$, then it will produce $q_1$ due to inertia or choose a best response to $q_{2}$, which must be no less than $q^{N}$. Applying this reasoning recursively, we conclude that if both firms produce $q$ in the initial state, then in each subsequent period, at least one firm's output is no less than $q^{N}$. Therefore, the dynamics finally converges to an absorbing set $\text{mon}(q', IM)$, with $q^{N}\leq q'\leq q\leq h(q^{N})$. 

Applying the \cite{freidlin1984random} algorithm, we conclude that for any $q'\notin \Gamma\cap[q^{N},h(q^{N})]$, $\text{mon}(q',\text{IM})$ is not LRE. 

\vspace{0.5em}
\noindent \textbf{Case 2}\quad Consider the case of $n > 2$. Before proceeding to a formal proof, we first analyze Example~\ref{example: a Cournot oligopoly} to build intuition. Note that $h(q^{N}) = h(15) = 25$. Applying the same reasoning as in \ref{proof of proposition 2}, we conclude that $\text{mon}(q', \text{IM})$ is LRE for all $q' = 15, 16, \dots, 25$. Furthermore, since $\ell(25) = \frac{5}{3}$, it follows that $\text{mon}(q', \text{IM})$ is LRE for all $q' = 2, 3, \dots, 25$. Applying this logic recursively, and using the facts that $h(2) = \frac{166}{3}$, $\ell(55) = 0$ (since $\Delta(55, 0) > 0$), and $h(0) = 60$, we ultimately conclude that $\text{mon}(q, \text{IM})$ is LRE for all $q = 0, 1, \dots, 60$.

To establish the proof, we define sequences $ \{a_n\} \subset\mathbb{R}_{+}$ and $ \{b_n\}\subset\mathbb{R}_{+}$ as follows: 
\[
a_1 = q^{N}, \quad a_{n+1} = \ell(h(a_n)), \quad \text{and} \quad b_n = h(a_n) \quad \text{for all } n \geq 1. \]
Because of \textit{(v)}, we have
\begin{equation}
\Delta(h(a_{n}),a_{n})=\Delta(a_{n},h(a_{n}))+\Delta(h(a_{n}),a_{n})>0.  
\label{eq: n > 2}
\end{equation}
Formula (\ref{eq: n > 2}) implies that, if $a_{n}>0$, then $\ell(h(a_{n}))<a_{n}$; otherwise, if $a_{n}=0$, then $\ell(h(a_{n}))=a_{n}$. That is, $\{a_{n}\}$ strictly decreases unless it reaches $0$. Furthermore, due to \textit{(vi)}, it follows that $\{b_{n}\}$ strictly increases unless it reaches $h(0)$.

For the decreasing sequence $\{a_{n}\}$, there are two possible scenarios. First, there exists $k\in \mathbb{N}$ such that $a_{k}=0$. Second, for any $n\in \mathbb{N}$, $a_{n}>0$. Note that the second scenario can be ruled out. Since $\{a_{n}\}$ is decreasing and bounded below by zero, it converges to some $a\geq0$, i.e., $\lim_{n \to \infty} a_n = a$. Furthermore, given that $h(\cdot)$ and $\ell(\cdot)$ are continuous (as established in \textit{(vi)}), we obtain \[
a=\lim_{n \to \infty} a_{n+1} =  \ell(h(\lim_{n \to \infty}a_n))=\ell(h(a)). 
\]
Note that $\ell(h(a))=a$ only when $a=0$. Hence, the increasing sequence $\{b_{n}\}$ converges to $h(0)$. Moreover, for any $\epsilon \in (0,\frac{h(0)}{2(n-1)})$, 
\[
\begin{aligned}
\Delta(h(0)-\epsilon, 0) 
&= -p((n-1)(h(0)-\epsilon))(h(0)-\epsilon) + c(h(0)-\epsilon) - c(0) \\
&= -p((n-1)(h(0)-\epsilon))(h(0)-\epsilon) + c(h(0)-\epsilon) - \big(c(h(0)) - p(h(0)) h(0)\big) \\
&= \big(p(h(0)) - p((n-1)(h(0)-\epsilon))\big) h(0) + \epsilon \big(p((n-1)(h(0)-\epsilon)) - c'(\xi^\epsilon)\big) \\
&\geq \big(p(h(0)) - p((n - 1.5)h(0))\big) h(0) + \epsilon \big(p((n - 1)h(0)) - c'(h(0))\big)
\end{aligned}
\] where $\xi^{\epsilon}\in [h(0)-\epsilon,h(0)]$. Note that $n\geq 3$ and $h(0)<Q_{max}$, hence it yields 
$$p(h(0))-p((n-1.5)h(0))>0.$$ 
We conclude that $\Delta(h(0)-\epsilon, 0)>0$ as long as $\epsilon$ is sufficiently close to zero, which contradicts with the assumption of the second scenario that $a_{n}>0$ for any $n$.

Provided that the finite strategy set $\Gamma$ is sufficiently fine, $\{a_n\}_{n=1}^{k}$ and $\{b_n\}_{n=1}^{k}$ are contained in $\Gamma$, where $a_k=0$ and $b_k=h(0)$. Applying the transition triggered by one action mistake recursively, we conclude that $\text{mon}(q,IM)$ is LRE for any $q\in\Gamma\cap[0, h(0)]$, when $n>2$. Moreover, it is evident that no further LRE exists. \hfill \qedsymbol
\newpage

\section{Appendix: Aggregative Games }\label{App: B}
\renewcommand{\thefigure}{B.\arabic{figure}}
\setcounter{figure}{0}
\renewcommand{\theequation}{B.\arabic{equation}}
\setcounter{equation}{0}

We now introduce symmetric aggregative games \citep{corchon1994comparative}, 
where agents choose actions (rather than firms choosing quantities) from a finite strategy set $S \subseteq \mathbb{R}$, forming a lattice.

\begin{defn}[\textbf{\textit{Aggregative games}}]
\label{def: aggregative games}
A symmetric \textit{aggregative game} with aggregate $g$ is defined as the tuple $G := (N, S, \tilde\pi, g)$, where $g: S^N \to T$ is a real-valued function that is symmetric and monotone increasing (referred to as the \textit{aggregator}), and $\tilde\pi: S \times T \to \mathbb{R}$ is a real-valued function such that the individual payoff functions are given by
$\pi_i(\bm s):=\tilde\pi\bigl(s_i,\;g(\bm s)\bigr)$, for all action profile $\bm s=(s_1,\dots,s_n)\in S^n,\ i=1,\dots,n$.
\end{defn}

We focus on aggregative games with strict quasi-submodularity. Prominent examples include Cournot oligopolies, the tragedy of the commons, rent-seeking games, and common-pool resource games. Refer to \cite{schipper2004submodularity} and \cite{alos2005evolutionary} for more examples.

\begin{defn}[\textbf{\textit{Quasi-submodularity}}]
An aggregative game $G = (N, S, \tilde\pi, g)$ is \textit{strictly quasi-submodular} if $\tilde\pi$ satisfies the following conditions: for any $s^{(1)} < s^{(2)} \in S$ and $t^{(1)} < t^{(2)} \in T$,
\[
\begin{aligned}
\tilde\pi\big(s^{(2)}, t^{(1)}\big) \leq \  \tilde\pi\big(s^{(1)}, t^{(1)}\big) 
&\quad\text{implies}\quad 
\tilde\pi\big(s^{(2)}, t^{(2)}\big) < \  \tilde\pi\big(s^{(1)}, t^{(2)}\big), \\
\tilde\pi\big(s^{(2)}, t^{(2)}\big) \geq \  \tilde\pi\big(s^{(1)}, t^{(2)}\big) 
&\quad\text{implies}\quad 
\tilde\pi\big(s^{(2)}, t^{(1)}\big) >\  \tilde\pi\big(s^{(1)}, t^{(1)}\big).
\end{aligned}
\]
\end{defn}

Analogous to the concept of a Walrasian quantity in the Cournot oligopoly, we now define an \textit{aggregate-taking strategy} for symmetric aggregative games \citep{possajennikov2003evolutionary}.

\begin{defn}[\textbf{\textit{Aggregate-taking strategy}}]
Assume $G = (N, S, \tilde\pi, g)$ is a symmetric aggregative game. A strategy $s^{*}\in S$ is an aggregate-taking strategy ($ATS$) if 
\begin{align*}
s^{*} \in \arg\max_{s \in S} \tilde\pi\big(s,\, g(\vec{s^{*}})\big),
\end{align*}
where $\vec{{s^{*}}}$ denotes the monomorphic action profile in which all agents choose $s^{*}$.  
\end{defn}

With this definition at hand, Formula (\ref{eq: Walrasian quantity yields higher relative payoff}) in the main text can be generalized as follows.

\begin{lemma}
\label{lemma: ATS}
We assume that the symmetric aggregative game $G = (N, S, \tilde\pi, g)$ is strictly quasi-submodular. Let $s^*$ be an $ATS$ of $G$. For all $s' \neq s^*$ and $1 \leq m < n$, we have
\begin{equation*}
\tilde{\pi}\Big(s^*,\,g\big((\underbrace{s', \dots, s'}_{m}, \underbrace{s^*, \dots, s^*}_{n - m})\big)\Big) 
\;>\; 
\tilde{\pi}\Big(s',\, g\big((\underbrace{s', \dots, s'}_{m}, \underbrace{s^*, \dots, s^*}_{n - m})\big)\Big). 
\end{equation*}
\end{lemma}

In this section, we assume that game $G$ admits an \textit{ATS}.\footnote{Kakutani’s fixed point theorem provides sufficient conditions for the existence of an \textit{ATS}.} Lemma \ref{lemma: ATS} further shows that the \textit{ATS} is unique whenever $G$ is a strictly quasi-submodular aggregative game. As established by \cite{schipper2004submodularity}, it follows that when all agents follow imitation learning, the $ATS$ equilibrium in which all agents play the \textit{ATS} is the unique LRE, thereby generalizing the result of \cite{vega1997evolution}. We denote by $\text{mon}(ATS, IM)$ the set of monomorphic states in which all agents adopt $IM$ and play the \textit{ATS}.

Moreover, we assume that $G$ admits a unique pure strategy Nash equilibrium, which must be symmetric given the symmetry of $G$. Following the same reasoning as in Appendix \ref{proof of proposition 1}, and relying on quasi-submodularity, we conclude that if all agents adopt best response learning, the dynamics converges globally to this Nash equilibrium. We denote  the set of monomorphic states in which all agents use $BR$ and play the Nash strategy as $\text{mon}(Nash, BR)$. 

Building on the above analysis, the absorbing sets can be readily identified. We further impose the mild assumption that the Nash strategy \textit{Nash}
 differs from the \textit{ATS}, and present the results as follows. 

\begin{proposition}
\label{proposition: aggregative games}
The set of long-run equilibria is characterized as follows.
\begin{enumerate}[label=(\roman*)]
    \item If $\eta>1$, then
    \[Z'=\text{mon}(Nash,BR)\bigcup \text{mon}(ATS,IM).
    \]
   \item If $\eta = 1$,  then 
    \[
    Z' \supseteq \text{mon}(Nash, BR)\bigcup \text{mon}(Nash, IM)\bigcup\text{mon}(ATS, IM).
    \] 
\end{enumerate}
\end{proposition}

The logic underlying Proposition~\ref{proposition: aggregative games} parallels that in Appendix~\ref{proof of proposition 2}, and the proof is therefore omitted. These results highlight that, despite the relative advantage of imitators in strictly quasi-submodular aggregative games \citep{schipper2009imitators}, best-response learning can nonetheless persist in the long run. Moreover, when rule mistakes occur at the same rate as action mistakes, the \textit{ATS} equilibrium is no longer the unique prediction under imitation. We point out that, beyond the Nash and \textit{ATS} equilibria, a range of action profiles may also emerge under imitation learning in the long run, following reasoning similar to that in Lemma~\ref{lemma: transitions by action mistake} and Appendix~\ref{Proof of the Remark}.

\bibliography{refs}
\bibliographystyle{elsarticle-harv}

\end{document}